\documentclass[aps,prl,twocolumn,superscriptaddress]{revtex4}
\usepackage{graphicx}
\begin{document}

\title{Infinite-randomness critical point in the itinerant quantum antiferromagnet}

\author{Rastko Sknepnek}
\affiliation{Department of Physics, University of Missouri-Rolla, Rolla, MO 65409}
\affiliation{Institut f\"ur Physik, Technische Universit\"at, D-09107 Chemnitz, Germany }

\author{Thomas Vojta}
\affiliation{Department of Physics, University of Missouri-Rolla, Rolla, MO 65409}

\date{\today}

\begin{abstract}
We study the quantum phase transition in the three-dimensional disordered itinerant
antiferromagnet by Monte-Carlo simulations of the order-parameter field theory.
We find strong evidence for the transition being controlled by an infinite-randomness
fixed point: The dynamical scaling is activated, i.e., the logarithm of the energy
scales like a power of the length, implying a dynamical exponent of infinity.
The probability distribution
of the energy gaps is very broad and becomes broader with increasing system size,
even on a logarithmic scale.
\end{abstract}

%\keywords{}

\maketitle


Phase transitions in systems with quenched disorder are an important topic in
statistical physics. The critical behavior of systems with quenched disorder can
be divided into three classes, according to the behavior of the disorder under
coarse graining. In the first class, disorder decreases under coarse graining, and the
system becomes asymptotically homogeneous at large length scales. Technically, this
means the disorder is renormalization group irrelevant, and the transition is controlled
by a "clean" fixed point (FP). According to the Harris criterion \cite{Harris74} this happens if
the clean FP fulfills  $\nu\ge 2/d$, where $\nu$ is the correlation length critical
exponent and $d$ is the spatial dimensionality.
In this first class the macroscopic observables are self-averaging at the critical point,
i.e., the relative width of their probability distributions goes to zero in the thermodynamic
limit \cite{AharonyHarris96,WisemanDomany98}.

In the second class, the system remains inhomogeneous at all length scales with the
relative strength of the inhomogeneities approaching a finite value for large length scales.
These transitions are controlled by renormalization group FPs with finite disorder.
Macroscopic observables
are not self-averaging, the relative width of their probability distributions
approaches a size-independent constant \cite{AharonyHarris96,WisemanDomany98}.
Examples of critical points in the second class include the dilute three-dimensional Ising model,
classical spin glasses, and various other thermal critical points in disordered systems.

The third possibility occurs when the relative magnitude of the inhomogeneities increases
without limit under coarse graining. The corresponding renormalization group FPs
are called infinite-randomness FPs. At these FPs the  probability
distributions of macroscopic variables become very broad (on a logarithmic scale) with the
width increasing with system size. Consequently, averages will be often
dominated by rare events, e.g., spatial regions with atypical disorder configurations.
Infinite-randomness critical points have mainly been found for quantum
phase transitions since the disorder, being perfectly correlated in (imaginary) time direction,
has a stronger effect for quantum phase transitions than for thermal ones.
Examples include the random transverse-field
Ising and Potts chains \cite{Fisher9295,SenthilMajumdar96,YoungRieger96} and
the two-dimensional transverse-field Ising model \cite{Pich98,Motrunich00}.

\begin{figure}
\includegraphics[width=\columnwidth]{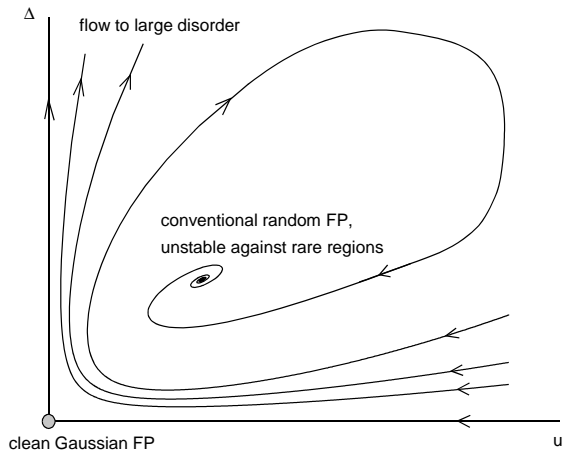}%
\caption{Schematic of the perturbative renormalization group flow.
         $\Delta$ is the disorder strength and $u$ is a measure of the quantum
         fluctuations}
\label{fig:flow}
\end{figure}

A natural question is, how general is the occurrence of infinite-randomness FPs in disordered
quantum systems.  One prototypical
and particularly controversial transition is the antiferromagnetic quantum phase transition of disordered itinerant
electrons in three dimensions which is believed to be important, e.g., for a variety of
heavy-fermion materials \cite{Andrade98}. It has been investigated by various methods
but no definite results could be achieved.
In the absence of disorder this transition is controlled by a Gaussian
FP with mean-field static critical exponents and a dynamical exponent of $z=2$ \cite{Hertz76}.
According to the Harris criterion \cite{Harris74} this clean FP is unstable with respect
to disorder, and the transition, if any, must be in class two or three of the above classification.
Within the conventional perturbative renormalization group \cite{BelitzKirkpatrick96}
one finds a finite-disorder FP suggesting that the transition belongs to
class two. However, the renormalization group flow diagram, Fig.\ \ref{fig:flow},
shows that a system with weak initial (bare) disorder is taken to large
disorder at intermediate stages of the renormalization before spiraling into the FP.
This casts serious doubts on the
validity of the perturbative approach. Indeed, by taking into account the effects of rare regions
it was later shown \cite{NVBK99} that the conventional FP is unstable and the
renormalization group flow is towards large disorder in all of the physical parameter space.
The ultimate fate of the transition, however, could not be determined within this approach.
Possible scenarios included a complete destruction of the transition, a conventional FP
 (i.e., class two) inaccessible by perturbative methods, or an infinite-randomness FP.

In this Letter we contribute to the solution of this puzzle by reporting results from large-scale computer
simulations of the antiferromagnetic
quantum phase transition of disordered itinerant electrons in three dimensions with uniaxial
(Ising) symmetry. We find strong evidence for a sharp phase transition which is controlled by an
infinite-randomness FP similar to those in the
random transverse field Ising models \cite{Fisher9295,SenthilMajumdar96,YoungRieger96,Pich98,Motrunich00}.
Specifically, we find that the probability distribution of the order parameter susceptibility $\chi$
(the inverse energy gap) becomes broader with increasing system size, even on a logarithmic scale.
The entire distribution scales with $\ln \chi \sim L^\psi$ with $\psi \approx 0.39$.
This implies a dynamical exponent of $z=\infty$.
The average gap and the average order parameter susceptibility follow stretched exponentials with
different powers. In the rest of the paper we outline our calculations and discuss the results in detail.

The simulation of the full disordered interacting electron problem is numerically very expensive.
Consequently, the system sizes are so severely restricted that an investigation of the
critical behavior is essentially impossible in three dimensions. Rather than simulating the full problem
we have therefore performed simulations of the Landau-Ginzburg-Wilson (LGW) theory  of the phase transition which
contains only the
long-wave-length, low-frequency fluctuations of the order parameter. This approach is valid close
to the transition as long as the (bare) disorder is weak so that local phenomena like
the formation of localized moments do not play an important role.

The LGW theory of the antiferromagnetic quantum phase transition of itinerant
electrons in $d$ dimensions is equivalent to a classical ferromagnet in $d+2$ dimensions with the
disorder being uncorrelated in the $d$ "space-like" dimensions but completely correlated in the
two "time-like" dimensions \cite{BelitzKirkpatrick96,BoyanovskyCardy82}. In the following we concentrate
on three spatial dimensions and uniaxial symmetry. For an efficient simulation
we remap the field theory to a lattice model. The quenched disorder is introduced via site dilution
in the "space-like" dimensions, i.e., the impurities are two-dimensional holes in the five-dimensional lattice.
The classical model Hamiltonian reads
\begin{eqnarray}
H= - \sum_{\langle {\bf r}{\bf \tau}, {\bf r'}{\bf \tau'\rangle}} \epsilon_{\bf r} \epsilon_{\bf r'} ~S({\bf r},{\bf \tau}) S({\bf r'},{\bf \tau'})~.
\label{eq:Hamiltonian}
\end{eqnarray}
Here ${\bf r}=(x,y,x)$ and ${\bf \tau}=(\tau_1,\tau_2)$ are the "space-like" and "time-like"
coordinates of lattice sites, respectively; and the sum runs over all pairs of nearest neighbors.
$S({\bf r},{\bf \tau})=\pm 1$ is an Ising spin, and
$\epsilon_{\bf r}$ is a quenched random variable with values 0 and 1 and an average of
$[\epsilon_{\bf r}]=p$ (the symbol $[\cdot]$ denotes the disorder average).
All the results in this Letter are for $p=0.8$ while the percolation
threshold for three-dimensional site percolation is $p=0.31$.  The rather weak disorder allows us
to observe the crossover from the clean Gaussian FP to the infinite-randomness
FP using the system size as a renormalization cutoff.
In the model
Hamiltonian the transition is tuned by changing the classical temperature $T_{cl}$ which is different from the
physical temperature of the quantum system; the latter is encoded in the length of the system in
time-direction.

For the Monte-Carlo simulations we have employed the Wolff cluster algorithm \cite{Wolff89}.
We have studied systems with linear sizes up to $L=29$ in space direction and
also $L_\tau=29$ in time direction (the largest system having 20 million sites).
We found that no more than 200 sweeps were required for equilibration.
Between 1000 and 5000 disorder configurations were considered, depending on system size with
at least 1000 measurement sweeps per disorder configuration.

To get an overview over the behavior of the system we have computed the average
Binder parameter
\begin{equation}
[g] = \left[ 1 - \frac {\langle M^4\rangle}{3\langle M^2\rangle^2} \right]~.
\end{equation}
Here $M$ is the magnetization and $\langle \cdot \rangle$ denotes the thermodynamic average
for a single sample. $[g]$ has the
expected finite-size scaling form
\begin{eqnarray}
 [g] &=& \tilde g(tL^{1/\nu}, L_\tau/L^z) \quad \textrm{conventional FP,}
 \label{eq:fss_conventional}\\{}
 [g] &=& \tilde g(tL^{1/\nu}, \ln L_\tau/L^\psi) \quad \textrm{infinite-randomness FP}.
 \label{fig:fss_infinite}
\end{eqnarray}
Here $t=(T_{cl}-T_c)/T_c$ is the distance from the critical point.
The dynamical scaling is of power-law type at a conventional FP but
activated at an infinite-randomness FP \cite{Fisher9295}.
As a result of its scale dimension being zero, $[g]$ is easily analyze:
provided $L_\tau$ is scaled appropriately with $L$,
a point in parameter space where $[g]$ is independent of the system size corresponds to
a renormalization group FP. In principle,
both $\nu$ and $z$ (or $\psi$) can be determined from the finite-size scaling of $[g]$ \cite{Pich98}.

\begin{figure}
\includegraphics[width=\columnwidth]{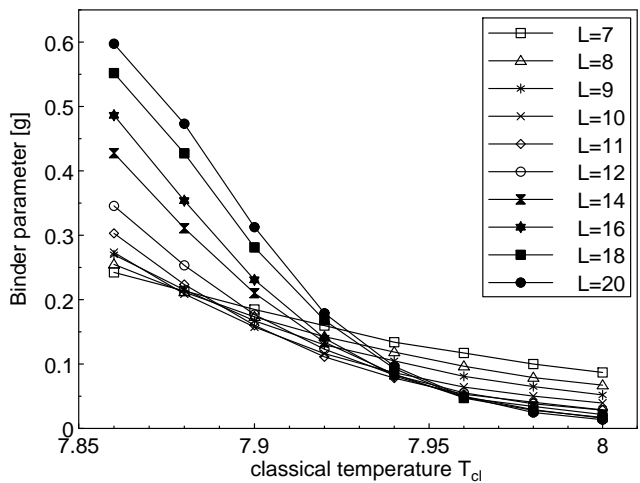}%
\caption{Average Binder parameter $[g]$ as a function of classical temperature $T_{cl}$ for
         different system sizes.}
\label{fig:bind_overview}
\end{figure}

In Fig.\ \ref{fig:bind_overview} we show the Binder parameter $[g]$ as a function of
the classical temperature $T_{cl}$ for different system sizes $L=L_{\tau}$.
For small
sizes, $L=7 \ldots 10$, the curves cross at $T_{cl}\approx 7.88$. However, for larger sizes this
crossing point becomes unstable and a new crossing emerges at $T_{cl}\approx 7.96$. A comparison
with the flow diagram in Fig.\ \ref{fig:flow} suggests that the former crossing point
corresponds to the clean Gaussian FP. (A system with small bare disorder first approaches
the Gaussian FP under renormalization before it goes to large disorder.)
To check this hypothesis we have performed a series
of calculations at $T_{cl}= 7.88$, independently varying $L$ and $L_\tau$.
The maximum of $[g]$ as a function of $L_\tau$ occurs at $L\approx L_\tau$ as expected
at the Gaussian FP. A standard finite-size scaling analysis
based on (\ref{eq:fss_conventional}) yields exponents of $\nu\approx 0.55$ and $z\approx 1.0$.
Since the range of available sizes at this unstable crossing is very limited the agreement
with the expected values $\nu_G=0.5$ and $z_G=1$ is surprisingly good \cite{z_footnote}.
We conclude that the crossing point at $T_{cl}\approx 7.88$ indeed corresponds to the clean Gaussian FP.

\begin{figure}
\includegraphics[width=\columnwidth]{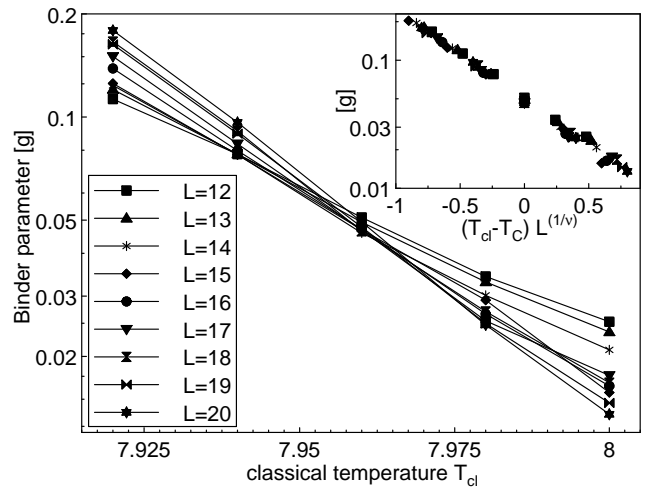}%
\caption{Log. plot of the Binder parameter $[g]$ close to the critical point.
         At $T=7.96$ the standard deviation is about one symbol size.
        Inset: Finite-size scaling plot using $\nu=1$.}
\label{fig:scaled_bind}
\end{figure}

We now turn to the crossing point at $T_{cl}\approx 7.96$ which we attribute to the
critical FP of the phase transition. We have again performed a series of calculations independently
varying $L$ and $L_\tau$. However, the dependence of $[g]$ on $L_\tau$ for fixed $L$
turned out to be very weak. (At $T_{cl}=7.96$ and $L=18$, $[g]$ remained constant at $g_c=0.048$
within our statistical error of about 6\% for all $L_\tau$ between 12 and 29.) This can be explained
by the fact that the critical value, $g_c=0.048$, is very small, which corresponds to weak average
correlations. This tends to reduce the influence of the boundary conditions and thus the sample shape
\cite{shape_footnote}. Consequently, finite-size scaling of $[g]$ is not an efficient method to
study the dynamical scaling of this critical point.
For all further calculations we have therefore used samples with $L=L_\tau$.
Fig.\ \ref{fig:scaled_bind} shows the average Binder parameter $[g]$ for systems with $L\ge 12$
in the vicinity of the critical point at $T_{cl}=7.96$.
The inset shows that the data scale reasonably well with a correlation length exponent of $\nu=1.0$.

\begin{figure}
\includegraphics[width=\columnwidth]{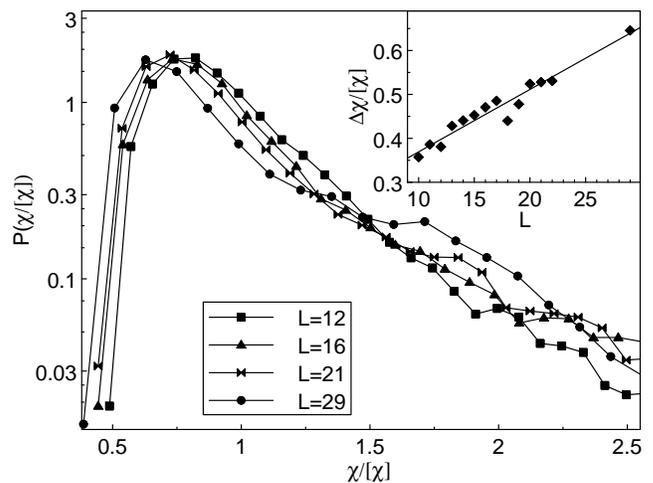}%
\caption{Distribution of the order parameter susceptibility $\chi/[\chi]$ at
         the critical point $T_{cl}=7.96$ for different sizes (at least 2300
         disorder configurations were used for the distributions).
         Inset: Dependence of $\Delta\chi/[\chi]$ on $L$. }
\label{fig:susc_dis}
\end{figure}

In order to further analyze the critical behavior we now consider the probability distribution
of the order parameter susceptibility $\chi$, i.e., the inverse energy gap.
At a conventional finite-disorder FP the distribution of $\chi/[\chi]$ should
be size-independent \cite{AharonyHarris96,WisemanDomany98}. In Fig.\ \ref{fig:susc_dis}
we show this distribution at the critical point, $T_{cl}=7.96$, for different system sizes.
Clearly the distribution becomes broader with increasing $L$. This can also be seen from
the inset which shows the relative widths of the distribution as a function of system size.
This suggests that the phase transition is controlled by an infinite-randomness FP
rather than a conventional one.

\begin{figure}
\includegraphics[width=\columnwidth]{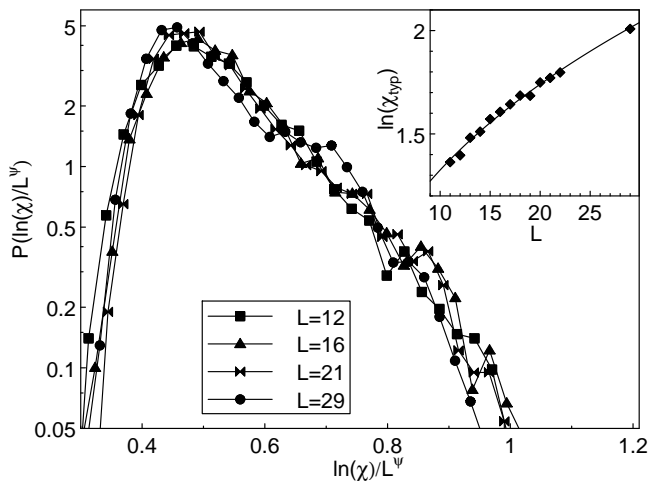}%
\caption{Distribution of $x=\ln \chi/L^\psi$ at $T_{cl}=7.96$ using $\psi=0.39$. Inset:
         Dependence of $\ln \chi_{typ}$ on L.
         The solid line is a power law fit with the same exponent $\psi=0.39$.}
\label{fig:fisherscaling}
\end{figure}

At an infinite-randomness FP we expect activated scaling \cite{Fisher9295},
$\ln \chi \sim L^\psi$. In Fig.\ \ref{fig:fisherscaling} we show a corresponding scaling
plot of the distribution of $\ln \chi$
The distribution
scales with an exponent of $\psi=0.39$. The inset shows the size-dependence of the logarithm of the
typical susceptibility, $\ln \chi_{typ}=[\ln \chi]$. The solid line is a power-law fit,
$\ln \chi_{typ} \sim L^\psi$, which gives the same exponent of $\psi=0.39$.

We now consider the {\em average} susceptibility $[\chi]$ and the {\em average} gap
$[1/\chi]$. Since  for large $L$ the distribution of $\ln\chi$ becomes very broad,
$[\chi]$ will be dominated by the large-($\ln\chi$) tail of the distribution.
For large $x=\ln\chi/L^\psi$ the distribution  in Fig.\ \ref{fig:fisherscaling} approximately
falls off as $e^{-c * x^2}$. For large $L$ the integral for $[\chi]$ can be treated in
saddle point approximation giving an asymptotic size dependence of  $\ln [\chi] \sim L^{2\psi}$.
The average order parameter
susceptibility increases faster with system size than the typical one.
Analogously, the average energy gap $[1/\chi]$ will be dominated by the
small-($\ln\chi$) tail of the distribution. However,  the accuracy
of our data is not sufficient to determine the functional form of this tail.
For system sizes $L=12$ to $29$ the susceptibility distribution is not sufficiently
broad yet for the above saddle-point argument to apply. In this size range we numerically obtain
the {\em effective} relations $\ln [\chi] \sim L^{0.45}$ and $-\ln [1/\chi] \sim L^{0.37}$.
While the exponent of $\ln [\chi]$ is larger than $\psi$, it is much smaller than
the asymptotic value of $2\psi =0.78$.

In summary, we have found strong evidence for an infinite-randomness FP
at the antiferromagnetic quantum phase transition of disordered itinerant electrons in three
dimensions. This FP has properties similar to those found in the random transverse-field
Ising model in one and two dimensions. We conclude with two remarks.
First, the systems we were able to simulate are rather modest in {\em linear} size
because the effective dimensionality of the LGW theory is very high.
Therefore all quantitative results for critical exponents should be understood as effective
rather than asymptotic values. Second, the results obtained here are for the case of uniaxial
symmetry. The qualitative features of the perturbative flow diagram in Fig.\ \ref{fig:flow}
including the runaway flow to large disorder, are the same for Ising, XY or Heisenberg symmetry.
This suggests that the transition is controlled by an infinite-randomness FP
for the XY and Heisenberg cases, too.
However, in Ref.\ \cite{Motrunich00} it was found that the random-singlet FP in localized
XY or Heisenberg antiferromagnets \cite{MDH79,BhattLee82,Fisher94}, another example of an infinite-randomness
FP, becomes unstable for $d>1$, suggesting that a continuous
order parameter symmetry tends to weaken infinite-randomness FPs. It thus remains unclear
whether the antiferromagnetic quantum phase transition of disordered itinerant electrons with
XY or Heisenberg symmetry is controlled by any of the know types of FPs or whether it
belongs to a new class.

%\begin{acknowledgments}
%\end{acknowledgments}

We gratefully acknowledge discussions with Dietrich Belitz, Ted Kirkpatrick
and Rajesh Narayanan. This work was supported in part by the German Research
Foundation under grant nos. Vo659/2, Vo659/3 and SFB 393/C2 and by the University
of Missouri Research Board. The bulk of the calculations was performed on the
CLIC parallel supercomputer of Chemnitz University of Technology, using a total
of about 6000 CPU days.



\begin{thebibliography}{99}
\bibitem{z_footnote} Note that the value of the dynamical exponent $z$ in the original quantum
         problem is twice a s large as that of our classical lattice model
         since we have mapped the time direction
         onto two additional space dimensions.
\bibitem{shape_footnote} This is in marked contrast to simulations of the random transverse-field Ising
         models \cite{Pich98} where the maximum $[g]$ at the critical point is approximately 0.55 in 1D and
         0.35 in 2D. Correspondingly, the dependence of $[g]$ on the shape is stronger and the dynamical
         exponent can be determined from a finite-size scaling analysis of $[g]$.
\end{thebibliography}
\end{document}